# THEMATIC ANALYSIS AND VISUALIZATION OF TEXTUAL CORPUS

Anja Habacha Chabi, Férihane Kboubi, Mohamed Ben Ahmed

RIADI Lab ENSI, University Compus of Manouba, Manouba, Tunisia
Anja.Habacha@ensi.rnu.tn, Ferihane.Kboubi@riadi.rnu.tn,
Mohamed.benAhmed@riadi.rnu.tn

## ABSTRACT

*The semantic analysis of documents is a domain of intense research at present. The works in this domain can take several directions and touch several levels of granularity. In the present work we are exactly interested in the thematic analysis of the textual documents. In our approach, we suggest studying the variation of the theme relevance within a text to identify the major theme and all the minor themes evoked in the text. This allows us at the second level of analysis to identify the relations of thematic associations in a textual corpus. Through the identification and the analysis of these association relations we suggest generating thematic paths allowing users, within the frame work of information search system, to explore the corpus according to their themes of interest and to discover new knowledge by navigating in the thematic association relations.*

## 1. INTRODUCTION

In the field of scientific and technical information management, thematic analysis of scientific documents is a still opened problem. Several works in the literature were interested in the thematic analysis [1][8][18][23][25]. However, we noticed that the process of thematic analysis is often confused by or reduced to the identification of themes. So it is interesting to start by presenting the difference between these two processes [22]. Indeed, although it partially consists of this last process, thematic analysis would not be reduced to it. As underlined in [3], "Thematic identification is the part of the thematic analysis task which allows the determination of the theme evoked in a textual unit". Thematic analysis can contain indeed several other processes. The specificity which seems to characterize the process of thematic analysis (by opposition to the identification of themes) lies mainly in the identification of the structure and the possible links between the various themes.

In the present work, we deal about thematic analysis. The remaining of this paper is organized as following. In the second section, we present a survey of the state of the art concerning thematic analysis. In the third section we describe the general principle of our approach of thematic analysis which is composed of two parts. The first one concerns term vector enriching and concepts identification. The second part concerns thematic extraction and analysis. Finally, in the fourth section, we present and discuss the obtained experimental results.

## 2. RELATED WORK

In the literature, the detection of theme changes is realized by two major classes of approaches. The first class of approaches is based on the identification of the linguistic markers in the text like "*about…*", "*it is a question of…*", "*as regarding…*" [7]. These approaches have the disadvantages to be language dependent and not to be always practicable since they require the existence of these markers inside the text.

The second class of approaches is based on the detection of the speech content changes [6] [9][24]. This second class of approaches presents the advantages to be independent of languages and always practicable. The principle is generally composed of three steps. In the first step, the document is divided into elementary blocks (windows of words, sentences, paragraphs). Secondly, the similarity between blocks is measured using often the notion of lexical cohesion. Thirdly, the thematic borders are identified by grouping the strongly similar elementary blocks. Various techniques were defined to measure the lexical cohesion between blocks of text: some of them are based on lexical measures of similarity [9][24], some others are based on the notion of lexical repetition [4], and some others are based on the semantic distance [6].

From our study of the state of the art, we noticed the lack of interest on studying the variation of theme relevance and the thematic associations in texts. We don't focus our attention in thematic segmentation of text or in textual summary extraction or generation. We are interested rather at identifying themes evoked in a textual document, and studying their respective pertinence which allow us annotating documents by one major theme and a set of minor themes. Based on this first thematic analysis process, we are also interested at studying thematic association relations in textual corpus.

From our point of view, through the thematic annotation, we aim to establish a global view of the textual corpus providing users with an information search system to localize their interest focus according to their themes of search. Indeed, such a module allows enriching document indexing process by identifying and mining thematic association relations and grouping together documents handling close themes. This facilitates users in finding similar documents to their document of interest.

### 3. GENERAL PRINCIPAL OF OUR THEMATIC ANALYSIS APPROACH

In this paper, we propose a new approach of thematic analysis of textual corpuses. The specificity of our approach is that it considers two levels: the document level (local analysis) and the corpus level (global analysis). As it is shown in the Figure 1, our method is based on seven steps. At the top of our thematic analysis process, there is a pretreatment step during which we eliminate from the documents all terms semantically insignificant. The remaining six steps are grouped in two phases: the conceptual annotation phase and the thematic annotation phase.

Our conceptual annotation method is composed of four steps: (1) relevant term extraction, (2) cooccurrence network construction, (3) term vector enriching and (4) concept identification. The result of these steps is the construction of a conceptual network representing the association relations between concepts and will be used as input for the thematic analysis.

Our thematic analysis method is composed of three steps: (1) theme detection and pertinence analysis, (2) identification of thematic association relations and (3) the generation and graphic visualization of thematic paths.

### 3.1. Conceptual analysis

The specificity of our conceptual analysis approach of textual documents is that it is based on a combination of an automatic summarizer and a semantic indexer which will be afterward enriched by the identification and the analysis of the cooccurrence relations.

### 3.1.1. Relevant Term Extraction

This module is based on the combination of an automatic summarizer [2] (using the thematic segmentation algorithm TextTiling [9] and a semantic indexer LSI [5]. The interest of using an automatic summarizer is to remediate to the problem of the voluminous size of documents and corpuses which could lead to the heaviness of treatments. An automatic summarizer allows reducing the space of the indexation terms without risking eliminating relevant terms. The interest behind using a semantic indexer is to be able to weight terms extracted by the automatic summarizer according to their density in the documents [14].

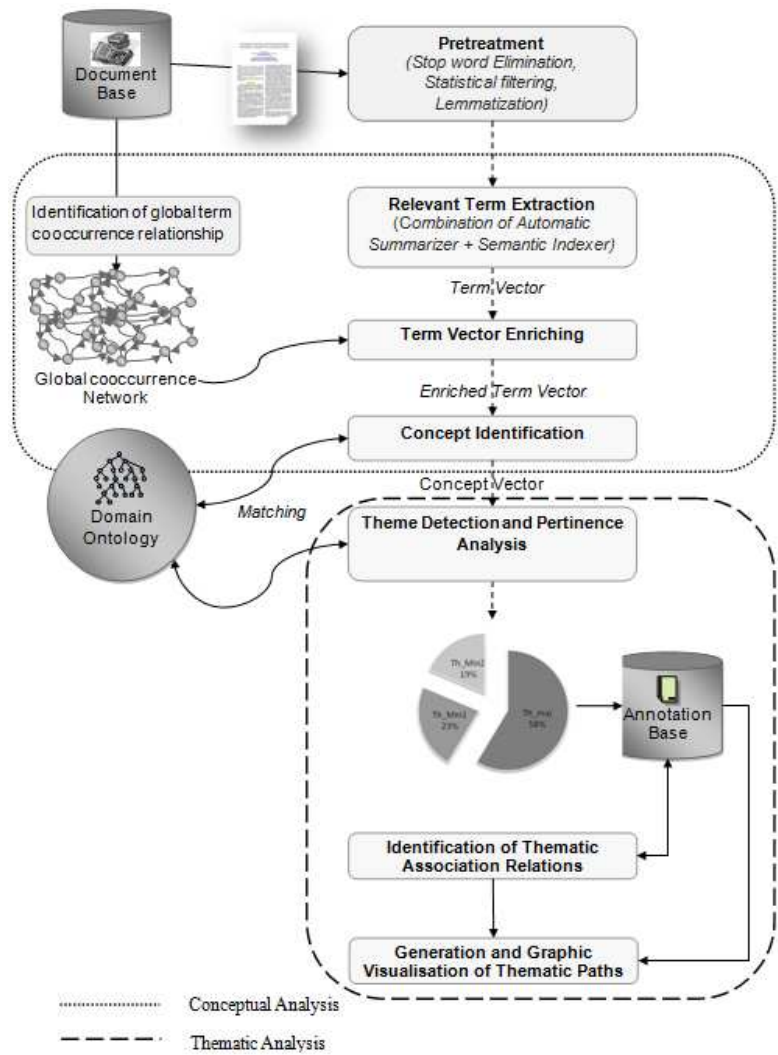

Figure 1. Thematic annotation of textual documents (Th_maj: Major Theme, Th_Min: Minor Theme).

### 3.1.2. Cooccurrence Network Construction

A cooccurrence network is a network of terms where every node represents a term and an edge between two nodes represents the relation of cooccurrence between both concerned terms. An association between two terms $t_1$ and $t_2$ is noted as following: $t_1 \rightarrow t_2$ it means that if $t_1$ appears then $t_2$ appears also in a window of size $L$. We weighted every edge by the confidence degree of the association rule of the two terms which it connects.

$$conf(t_1 \to t_2) = \frac{|t_1 \wedge t_2|}{|t_1|} \in [0,1] \qquad (1)$$

With $|t_i|$ is the number of documents containing the term $t_i$.

This step is independent of the other steps and could be realized at the beginning of the whole process.

At this level, it is important to notice that thematic analysis methods based only on term extraction are subjected to several problems in particular to the ambiguity errors. According to the context, a term can have several different senses. The term "*Network*" can indicate "*Semantic Network*", "*Computer Network*" or "*Neural Network*". To remediate to this problem we propose two solutions: explore the assoiciation relations between terms through the construction and the analysis of cooccurrence network [13] and identify semantic relations through the use of domain ontology to identify the corresponding concepts.

### 3.1.3. Enriching Term Vectors by Mining Cooccurrence Networks

The enrichment is made by two processes: (1) the *identification of compound terms* by the analysis of the cooccurrence of the simple terms at the level of the document; (2) the *inference of new terms* by adding to the descriptive vector the terms strongly associated to the term composing it.

In the ***first process***, we assume that the compound terms are more semantically significant than the simple terms [10][17]. As an example, the simple terms "*language*" and "*program*" taken separately can have meanings very different from the compound term "*programming language*".

Our process of term extraction based on the combination of automatic summarizer and semantic indexer cannot guarantee the extraction of compound terms. So, we propose to analyze the cooccurrence association relations of the simple terms in the document to identify the compound terms.

Our method begins with the construction of the coccurrence network of the document. Afterward, we identify the strong cooccurrence relations which could represent the compound terms. The compound terms obtained by this process are added to the vector of the descriptive terms of the document. The general principle of this process is summarized by the following algorithm:

---

***Let*** V be the term vector of the document D
***if*** $conf(t_1 \to t_2) = 1$
***then*** $t_3$:=new compound-term
    $t_3 \leftarrow t_1 + t_2$
    Add $t_3$ to V

---

In the ***second process***, the idea is to enrich the document vector through the inference of new terms strongly associated to those of the document descriptive vector. The deduced terms are inferred from the global cooccurrence network and may not belong to the initial document.

The association relations which we used in this second process are extracted from the global cooccurrence network which is built from a set of representative documents of the corpus.

The benefits behind the enriching of document's vectors by information deducted from the global cooccurrence network lies in the fact that we noticed that the association relations between terms can represent implicit semantic relations (like it is shown in Figure 2), such as:

- The synonymy (in the same paragraph one concept can be referenced by several different terms to avoid the redundancy). In the Figure 2, the term "*Imperfect*" is associated to "*Probable*", "*Possibl*" and "*Uncertainty*" and this last one is associated with "*Imprecis*".

- "*is-a*" : "*Mobil Agent*" and "*Agent*",

- "*is part of*": "*Pixel*" and "*Imag*".

- Strong dependency: ("*Mobil Agent*" and "*Network*"), ("Imag" and "Treatement").

Every term of the global cooccurrence network is described by a vector containing the terms which are strongly associated to it. Mining these vectors allows the consideration of the implicit semantic relations between the terms for the enrichment of the representations of documents.

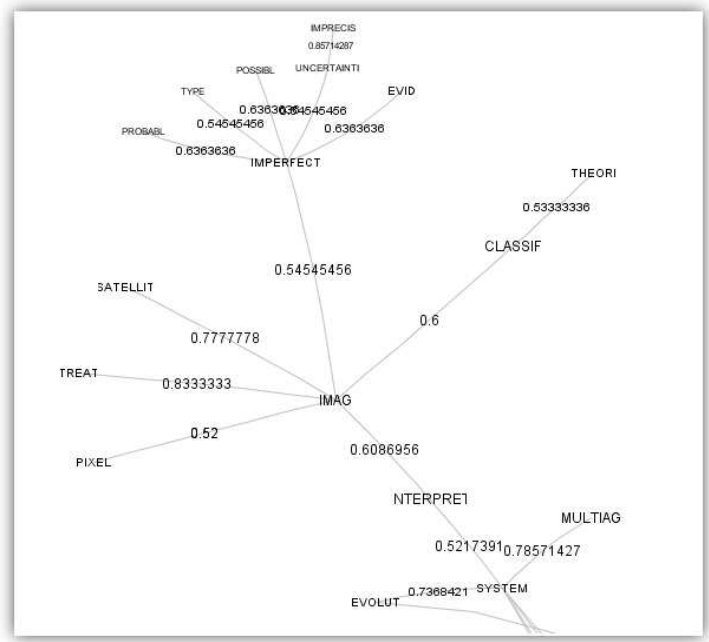

Figure 2. An extract of the global cooccurrence network: Implicit semantic relations

### 3.1.4. Concept Identification

The principle of our method of concept identification is based on domain ontology. It consists of identifying the concepts denoted by one or many key-term(s) of the document. Several cases are then possible. The simplest case is when the term extracted from the text allows referencing only a single concept $c$ (this is the case of the majority of the compound-terms). In that case, we retain as result the concept $c$. This concept is then added to the list of the validated concepts.

However, in some cases (especially when the term is simple, composed only of one word), we can have an ambiguity situation if the term could be used to denote two or more different concepts. In these cases, a mechanism of disambiguation becomes necessary. To deal with ambiguity situations, we suggest retaining the semantically closest concept to the key-concepts already identified by the other terms of the vector.

To estimate the proximity of concepts we used conceptual similarity measures. In the literature several measures were proposed. We distinguish in the first place, the measures based on the hierarchical relations such as the measure of Rada [20], the measure of Leacock [15] and the measure of Wu and Palmer [26]; secondly the measures based on the informative contents such as the measure of Resnik [21] and the Lin measure [16]; thirdly the hybrid measures such as of Jiang-Conrath [12] and fourthly the measures integrating the non taxonomic relations such as the measure of Hirst-St. Onge [11] and the measure of Hernandez [10].

In our work, we chose to test two measures using different principles, namely:

- Wu and Palmer measure which is based on counting the number of edges between the concepts [26],

- and Resnik measure which is based on information content [21].

Our desambiguisation method consists of calculating the sum of the similarity of every candidate concept (*cc*) with all other already identified and validated concepts (*cv*) by using the following formula:

$$score(cc_i) = \sum_{j=1}^{ncv} Simil(cc_i, cv_j) \qquad (2)$$

Then we retain the candidate concept who presents the maximal sum of similarity. This concept is added to the list of the validated key concepts.

### 3.2. Thematic analysis

Our vision of thematic annotation regroups two levels: local level (document) and global level (corpus). In the first level, our approach consists of identifying themes treated in a document. We think that a document is produced to describe a subject belonging to a specific theme. This theme forms the general theme of the document. Consequently, we consider that each document is characterized by one single "*major theme*" and that several other themes can be evoked in the document without being the principal goal of the document. These "*minor themes*" have an important role for the annotation of the document since they reflect the domain of association of the major theme. So we consider that a document is characterized by a *major theme* and by a network of *minor themes* (classified by importance) which denotes the variation of the themes within the same document.

In the corpus level, our vision of thematic analysis consists of giving a description of the thematic composition of the corpus according to the number of documents dealing with every theme. The thematic description of the corpus also highlights the multi-theme character of documents by giving a global view of the association relations between the themes.

### 3.2.1. Theme Pertinence Analysis

The analysis of the thematic variations within a document allows identifying the multi-theme character of a document and thematic associations in a document. Our method of theme identification is based on using concept vector instead of term vectors. For theme detection, we match the concepts extracted from the document, with domain ontology. A theme is defined by a cup of the domain ontology defining the concepts belonging to this theme as it is shown by Figure 3.

The principle of our theme identification method consists of parsing the domain ontology, for each relevant concept of the document, in order to identify the corresponding theme. If the obtained theme doesn't belong to the detected theme list then it is added to this list.

Matching the concepts of the document with the domain ontology can lead to two possible cases. In the first place, all the concepts of the document belong to the same theme. In that case, this theme will be considered as the main and unique theme of the document.

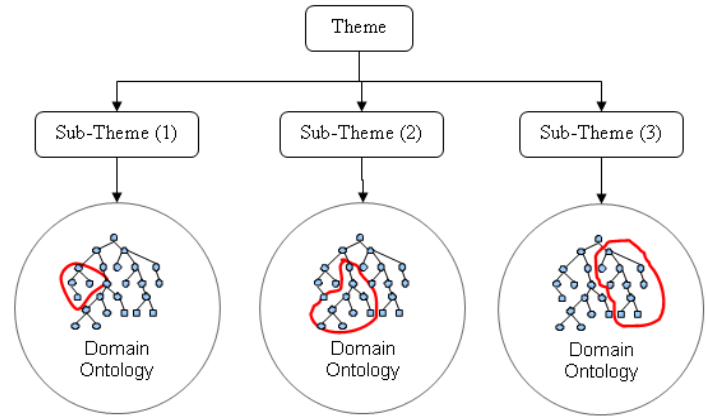

Figure 3. Architecture of ontology of themes

Secondly, the concepts of the document belong to two or several themes. In that case, we analyze the relevance of the themes inside the document according to the number of concepts of the document belonging to every theme. We measure the relevance of a theme according to the number and weights of the key concepts of the document which references it. The relevance of a theme $th_x$ is calculated by the following formula:

$$pertin(Th_x) = \frac{\sum_{i=1}^{nbConcepts_{Thx}} weight(c_i)}{nbConcepts} \quad (3)$$

Where $nbConcepts_{Thx}$ represent the number of key concepts relative to the theme $Th_x$, $nbConcepts$ represent the total number of key concepts of the considered document and $weight(c_i)$ represents the pertinence of the concept $c_i$. Every concept is weighted according to the number of key terms which denotes it by the following formula [13]:

$$weight(c_i) = \frac{\sum_{i=1}^{nbTerm_{ci}} weight(KeyTerm_{ci})}{nbKeyTerm} \quad (4)$$

Where $nbTerm_{cj}$ represents the number of key terms denoting the concept $c_j$, and $nbKeyTerm$ represent the total number of key terms.

We consider as major theme, the theme having the highest relevance, and as minor themes all other themes identified in the document. So the thematic annotation $TA$ of a document $d_i$ can be formalized by:

$$TA_{di} = \left((Th_{Maj}, pertin(Th_{Maj})), (Th_{\min}, pertin(Th_{\min}))^*\right) \quad (5)$$

Where $(Th_{Maj}, pertin(Th_{Maj}))$ represent respectively the label and the pertinence of the major theme of the document $d_i$; and $(Th_{\min}, pertin(Th_{\min}))^*$ represent the set of the label of minor themes with their respective pertinences as regards to the document $d_i$.

The general principle of our method of theme detection and weighting is described by the following algorithm:

```
if |listThemes|=0
then "No Theme is identified"
else if |listThemes|=1
     then Theme←listThemes.element(1)
 else
 for each theme of listThemes
  for each key-concept cᵢ belonging to thₓ
     pertin(thₓ)←pertin(thₓ) + weight(cᵢ)
  end for
        pertin(thₓ)←pertin(thₓ)/Theme_number
 end for
MajorTheme←find_Theme_MaxWeight
MinorThemeList←ThemeList-MajorTheme
end if
end if
```

### 3.2.2. Identification of Thematic Association

The thematic analysis of the corpus aims at determining the thematic distribution of documents in the corpus. It allows, on one hand, giving a global view of the corpus according to the number of documents belonging to every theme. Themes are weighted according to the number of documents on which they are tackled. The global weight of a theme represents the portion of documents of the corpus dealing with this theme. This weight is calculated as following:

$$weight(Th_x) = \frac{nbDoc(Th_x)}{nbDocTot} \quad (6)$$

Where *nbDoc(Th$_x$)* represents the number of documents dealing with the theme $Th_x$ as well as major theme or as minor theme; and *nbDocTot* represents the total number of document in the corpus.

On the other hand, it also allows studying the association relations between the themes. The association relation between two themes $Th_1$ and $Th_2$ is noted as following: $Th_1 \xleftrightarrow{AD(Th_1, Th_2)} Th_2$.

Where $AD(Th_1, Th_2)$ represents the degree of association between two themes. It is expressed according to the number of documents in which these two themes are approached together. This degree is measured by the following formula:

$$AD(Th_1, Th_2) = \frac{Doc(Th_1, Th_2)}{Doc(Th_1) + Doc(Th_2) - Doc(Th_1, Th_2)} \quad (7)$$

Where *Doc(Th$_i$)* represents the number of documents dealing with theme $Th_i$.

The analysis of the thematic relations is made by calculating the matrix of the associations of themes; where lines and columns represent the themes of the considered domain (Table 1). A null value means that the theme represented by the line and the theme represented by the column are not tackled together in any document of the corpus. If a value is equal to 1, this means that the theme of the line and the theme of the column always appear together in the documents of the corpus. If $Th_{line}$ appears in a document $d$ then $Th_{column}$ appears also in this document.

Table 1. Matrix of thematic association.

|        | $Th_1$ | $Th_2$ | $Th_3$ | … | $Th_n$ |
|--------|--------|--------|--------|---|--------|
| $Th_1$ | -      | 0.4    | 0.8    | … | 0.1    |
| $Th_2$ | 0.4    | -      | 0.3    | … | 0.6    |
| $Th_3$ | 0.8    | 0.3    | -      | … | 0.2    |
| …      | …      | …      | …      | … | …      |
| $Th_n$ | 0.1    | 0.6    | 0.2    | … | -      |

Once built this matrix allows identifying association relations between the themes of the domain. The periodic construction and analysis of this type of matrices allows detecting the trends and the evolutions of association between the themes.

The Figure 4 represents a formalization of the thematic annotation results of the corpus.

```
<!-- Thematic Annotation -->
<!ELEMENT thematicAnnotation(Stheme+,AssocTheme*)>
<!ELEMENT Stheme EMPTY>
<!ATTLIST Stheme
LAB CDATA
WEIGHT CDATA
>
<!ELEMENT AssocTheme EMPTY>
<!ATTLIST AssocTheme
theme1 CDATA
theme2 CDATA
WEIGHT CDATA
>
```

Figure 4. Extract from the DTD of thematic annotation of textual corpus

**3.2.3. Generation and Graphic Visualization of Thematic Paths**

The utility of a process of thematic analysis was often associated to the processes of document classification [25]; [23] and document summarization process [3].

In our work the main objective through the implementation of a process of thematic analysis is in the first place the enrichment of the document indexing process (a document is not indexed only by the set of its key terms but by its major theme and the set of its minor themes); and secondly the exploitation of these thematic knowledge on documents for the construction of graphic representation visualizing semantics contained in the textual documents.

Within the framework of Information Search System, this will allow the users to have a global view of the thematic composition of the text corpora and the various existing thematic associations. This would allow the users to discover new knowledge close to their initial domains of interest. Through the graphic visualization of the thematic annotations we also intend to propose to the users a means of navigation in the textual corpus by the generation of thematic paths.

For the visualization of the thematic associations we used the node-link representation paradigm. This paradigm allows us to build visual representations of the thematic networks through graphs where every node corresponds to a theme and every edge between two nodes corresponds to the association relation between both themes. The thematic visualization of networks by a node-link representation (in graph) leads to the identification of the thematic paths and facilitates the navigation in the indexed textual space.

The identification of thematic paths is based on the results of the previous steps in particular the step of identification of thematic association relations. Generally, the thematic navigation consists of a path characterized by a compromise between on one hand, the expectations of the reader (subjective constituent) and, on the other hand, the semiotic indications present in the text.

In our research, the subjective constituent is materialized by the interest of the researcher to certain themes (that he privileges) and in the goals which he wishes to reach during his analysis. So, the researcher can, for example choose to explore a precise theme of the corpus which interests him. The fact of offering to the user a global vision on the association relations of its theme of interest, allow him to discover new knowledge. This is possible by allowing the user to discover themes related to his initial search interest. This allows every user to build its thematic path by navigating and exploring the thematic association relations.

As shown in Figure 5 the user can navigate from one theme to another and for every theme $Th_i$, he can visualize the thematic association relations of $Th_i$ with their respective association degree $AD(i,j)$. This process can be repeated again infinitely up to the closure of the path.

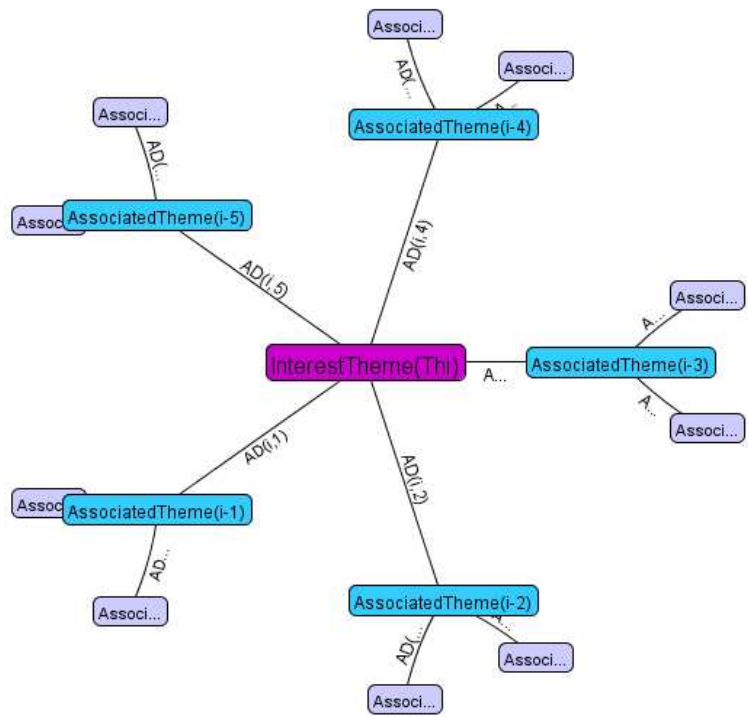

Figure 5. Principle of Thematic Association Relation Visualization

# 4. EXPERIMENTATION

The evaluations are realized on a test base constituted by a set of documents representing the publications of our laboratory during the three years 2006, 2007 and 2008. These documents are distributed in three major themes: Artificial Intelligence, Security and Information Systems. A pretreatment step is applied to all documents of the corpus. During which we remove the stop words. Then we make a statistical filtering to eliminate terms having a frequency lower than a threshold. Indeed, infrequent terms are the most numerous; consequently the number of indexation terms can increase in a very important way with the size of the corpus. The elimination of these terms is justified especially if we know that most of them are not representative of the document content and are not discriminating for its description. Finally, we proceed to the lemmatization of the remaining terms using the Porter algorithm [19]. The lemmatization allows to group together in a single attribute the multiple morphological forms of words which have a common semantics. In the remaining of this section, we start by presenting an experimentation result of our local thematic analysis method, and then we present and discuss the experimentation results of our method of global thematic association analysis.

Figure 6 illustrates an example of thematic annotation of a document from our corpus. In this figure, we present:

– The *key terms* of the document (extracted by the combination of TextTiling and LSI and enriched by the global cooccurrence network). Every term is accompanied by its respective weight.

– The *local cooccurrence network* of the key-terms. In this network each node represents a term and each edge represents the cooccurrence relation between the two terms. The labels of edges represent the confidence degree of the association relations.

– The *key concepts* identified by the conceptual annotation step.

– The *themes and the sub-themes* identified by the thematic annotation step. In this example, the major theme is "*Security*"; and there is only one minor theme which is "*Artificial Intelligence*".

– The *subtheme pertinence pie chart*. In this example "*Cryptography*" is the major subtheme, while the two other subthemes "*Network Security*" and "*Application and Expert System*" are both minor themes and have the same pertinence.

Figure 6 represents a summary of the result of document annotation task. It represents a semantic summary of the document which has a double utility in an information search system concerning both the document indexing process and the result visualization strategy. As regards to the indexing process, documents will not only be indexed by terms but also by associations of concepts and themes. As regards to the visualization strategy this allows to create semantic visualizations of results. The user will not any more need to visualize and to read the complete document to have an idea of its content. The summary of annotation gives him a clear idea about its content and allows him to decide more quickly on its relevance with regard to its need.

For the experimentation of our global thematic analysis method of the corpus, we begin by presenting the results of document thematic distribution analysis. Table 2 represents an extract of the thematic composition of our corpus where themes are arranged in the descending order of their weights. Figure 7 illustrates an extract of the XML file representing the thematic composition of the corpus. For every theme we mention its label with the attribute *LAB* and its weight with the attribute *WEIGHT*.

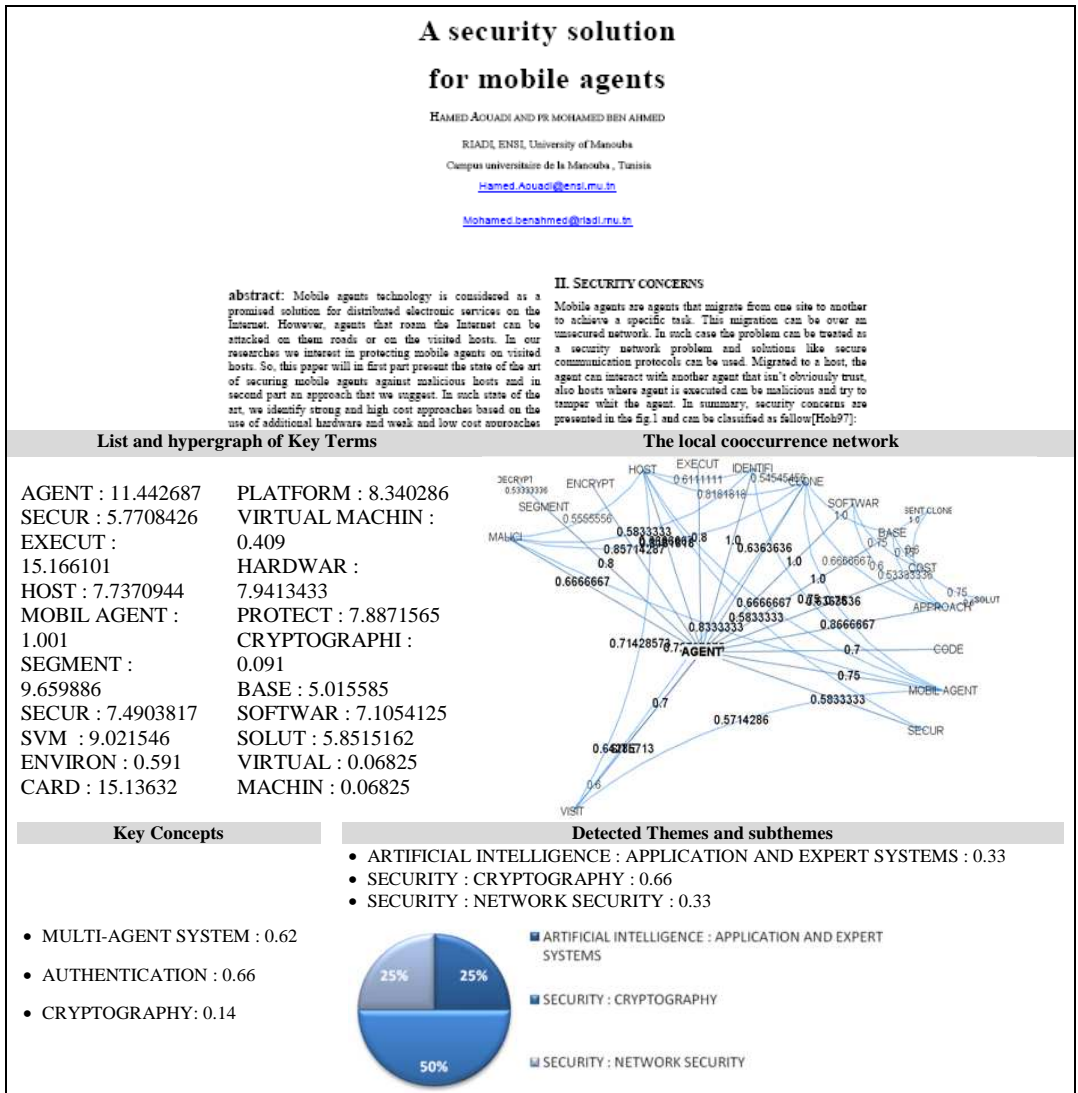

Figure 6. Example of thematic annotation result of a document [27].

Theme's weight is calculated according to the number of documents where this theme is approached as major theme or as minor theme. This figure shows that the subtheme "*Application and Expert Systems*" has a weight approximately equal to 0.5238. This means that 52.38 % of the documents of the corpus tackled this subtheme.

Table 2. Thematic composition of the textual corpus.

| Theme LAB | WEIGHT |
| --- | --- |
| Knowledge Representation Formalisms and Methods | 0.642 |
| Application and Expert Systems | 0.523 |
| Natural Language Processing | 0.309 |
| Information Storage and retrieval | 0.261 |
| Database Management | 0.214 |
| Network Security | 0.166 |
| Watermarking | 0.166 |
| Learning | 0.071 |
| Problem Solving, Control Methods and Search | 0.071 |
| Cryptography | 0.071 |

```
<Stheme LAB='KNOWLEDGE REPRESENTATION FORMALISMS AND METHODS'  WEIGHT ='0.64285713'/>
<Stheme LAB='APPLICATION AND EXPERT SYSTEMS'                    WEIGHT ='0.52380955'/>
<Stheme LAB='NATURAL LANGUAGE PROCESSING'                       WEIGHT ='0.30952382'/>
<Stheme LAB='INFORMATION STORAGE AND RETRIEVAL'                 WEIGHT ='0.26190478'/>
<Stheme LAB='DATABASE MANAGEMENT'                                WEIGHT ='0.21428572'/>
<Stheme LAB='NETWORK SECURITY'                                   WEIGHT ='0.16666667'/>
<Stheme LAB='WATERMARKING'                                       WEIGHT ='0.16666667'/>
<Stheme LAB='LEARNING'                                           WEIGHT ='0.071428575'/>
<Stheme LAB='PROBLEM SOLVING, CONTROL METHODS AND SEARCH'        WEIGHT ='0.071428575'/>
<Stheme LAB='CRYPTOGRAPHY'                                       WEIGHT ='0.071428575'/>
```

Figure 7. Extract from the XML file resulting from the analysis of thematic corpus composition.

Table 3 and Figure 8 represents the association relations between the themes where the attributes *theme1* and *theme2* represent both themes in relation and *WEIGHT* represents the degree of association between both themes. In Table 3 these relations are arranged in the descending order of their degree of associations.

Table 3.  Examples of theme association relations.

| Theme1 | Theme2 | WEIGHT |
|---|---|---|
| Application and Expert Systems | Knowledge Representation Formalisms and Methods | 0.441 |
| | Natural Language Processing | 0.206 |
| | Information Storage and Retrieval | 0.178 |
| | Database Management | 0.148 |
| | Network Security | 0.160 |
| | Problem Solving, Control Methods and Search | 0.136 |
| | Cryptography | 0.086 |
| | Watermarking | 0.074 |
| | Learning | 0.041 |

```
<AssocTheme theme1 ='APPLICATION AND EXPERT SYSTEMS'
            theme2 ='KNOWLEDGE REPRESENTATION FORMALISMS AND METHODS'
            WEIGHT ='0.44117647'/>

<AssocTheme theme1 ='DATABASE MANAGEMENT'
            theme2 ='INFORMATION STORAGE AND RETRIEVAL'
            WEIGHT ='0.42857143'/>
```

Figure 8. Extract of the XML file resulting from the thematic association analysis.

To make the understanding of the theme association relations easier for the users, we proposed to visualize graphically the results using the hypergraph visualisation tool applet (Available on line at http://hypergraph.sourceforge.net/). For each theme we constructed its association graph. The central node of every graph represents the theme of interest and the peripheral nodes represent the themes associated to the central theme. The labels of edges represent degrees of association between the themes. Figure 9 represents an example of theme association graphs concerning respectivelly the subthemes "*Application and Expert Systems*" and "*Knowledge Representation Formalisms and Methods*".

From the first association graph, we notice that the subtheme "*Natural Language Processing*" is associated to the subtheme "*Application and Expert Systems*" with a degree of association equal to 0.2068. This means that 20.68 % of the documents of the corpus which tackled the subtheme "*Application and Expert Systems*" tackled also the subtheme "*Natural Language Processing*".

Thematic analysis and visualization of textual documents allow the users of an information search systems to observe the dynamic evolution aspect of themes. This is important for many kinds of users since it offers to them a mean for observing the dynamics of apparitions, declines, developments of themes.

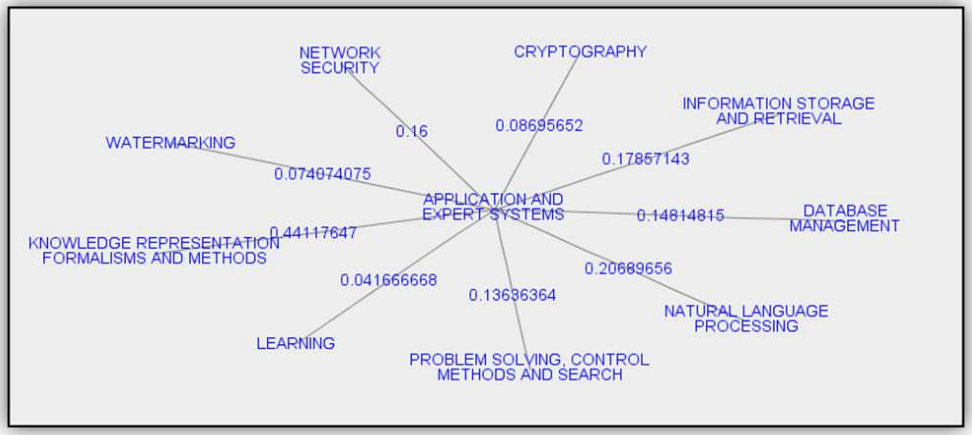

(a)

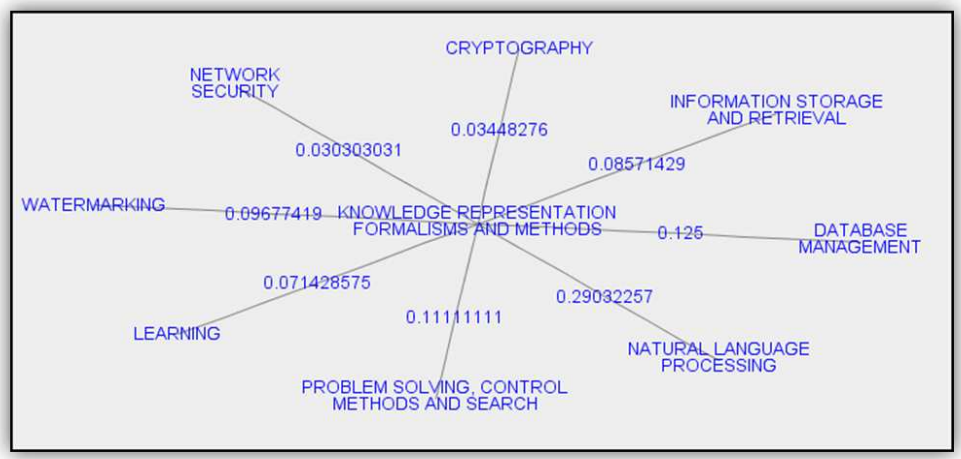

(b)

Figure 9. Hypergraph of association theme of the subtheme "Application and Expert Systems".

For a researcher this allows him to ameliorate his visibility of his own researches and to have a transversal vision since he could discover other works having different points of view. Similarly, for a team director this allows him to ameliorate his visibility about the whole work of his team and to be alerted about the significant evolutions of themes in order to detect the emergence of new trends and to be always up to date.

## 5. CONCLUSION

In this paper we proposed an approach of thematic annotation which deals with two levels: the local level of the document and the global level of the corpus. At the document level, thematic analysis consists, on one hand, at identifying the major theme and the set of minor themes of the document and, in the other hand, at studying their pertinences. Themes are weighted according to their importance and their variation within the same document. At the corpus level, thematic analysis consists, firstly, at giving a description of the thematic composition of the corpus according to the number of documents approaching every theme; and secondly, at identifying thematic associations in the corpus in order to highlight the multi-theme character of documents.

The originality of the proposed method is that it does not limited to the detection of the themes but also analyze their variations and their pertinences with regard to the document, and their global associations at the level of the corpus. We proposed also an hypergraph paradigm for the visualization of the thematic association relations. The utility of thematic association analysis concerns several tasks. Effectively, in information search systems, the thematic association analysis allows to index documents by associations of themes instead of single theme. This allows the users to make more complex searches.

From this work several perspectives can be considered. As a first future work we intend to integrate our approach of thematic analysis into an information search system and to improve the construction of graphic visualizations bringing to light the results of our process of thematic annotation.